\documentclass[onecolumn,prb]{revtex4}

\usepackage{epsfig,amssymb,amsmath}

\begin{document}

\title{Vortex structure in long Josephson junction with two inhomogeneities}

\author{O.Yu.Andreeva~$^{1}$}
\author{T.L.Boyadjiev~$^{2}$}
\author{Yu.M.Shukrinov~$^{1,3}$}

\address{
$^{1}$Tumen Thermal Networks OAO "TRGK", Tobolsk, 626150, Russia\\
$^{2}$Joint Institute for Nuclear Research, 141980 Dubna, Russia \\
$^{3}$Physical Technical Institute, Dushanbe, 734063, Tajikistan}

\begin{abstract}
A report of numerical experiment results on long Josephson junction with one and two
rectangular inhomogeneities in the barrier layer is presented. In case of one
inhomogeneity we demonstrate the existence of the asymmetric fluxon states. The
disappearance of mixed fluxon-antifluxon states when the position of inhomogeneity
shifted to the end of the junction is shown. In case with two inhomogeneities the change
of the amplitude of Josephson current through the inhomogeneity at the end of junction
makes strong effect on the stability of the fluxon states and  smoothes the maximums on
the dependence ``critical current - magnetic field''.
 Pacs: {05.45.+b}{74.50.+r}
74.40.+k. Keywords: long Josephson junction,  inhomogeneity, bifurcation, critical curve

\end{abstract}

\maketitle

In order to investigate a stability of fluxon states in inhomogeneous in-line Josephson
junction (JJ), we solve the following  non-linear eigenvalue problem
\begin{subequations} \label{inline}
    \begin{gather}
        -\varphi_{xx} + j_C (x)\sin \varphi  = 0,\\
        \varphi_x(0) = h_e - \varkappa_l\,L\gamma,\;\varphi_x(L) = h_e +
        \varkappa_r\,L\gamma,\\
  -\psi_{xx} + \psi\,j_C(x)\,\cos\varphi = \lambda \,\psi,\\
         \psi_x(0) = 0,\;\psi_x(l) = 0,\; \int\limits_0^L \psi^2(x)\,d x - 1 = 0,
    \end{gather}
\end{subequations}
with respect to the triplet $\left\{\varphi(x), \psi(x), h_e\right\}$. Here $\varphi(x)$
represents the static magnetic distribution in JJ, $L$ is junction's length, $h_e$ ---
external magnetic field, $\varkappa_l+\varkappa_r = 1$.

The  inhomogeneity in the form of the narrow  rectangular well is characterized by its
width $\Delta < L$, localization $\zeta \in [\Delta/2, L-\Delta/2]$ and portion of
Josephson current $\kappa$ through it. An existence of the inhomogeneity leads to the
local change of the Josephson current, which is equal to $ j_C (x) = 1 + \kappa$ inside
of the inhomogeneity and $j_C (x) = 1$ outside of it. At $\kappa > 0$ we have shunt, at
$\kappa \in [-1,0)$ --- microresistor. The value $\kappa = 0$ corresponds to the
homogeneous junction. A minimal eigenvalue of the Sturm-Liouville problem allow to make
a conclusion about the stability of the  $\varphi(x)$ (see details in \cite{tlb_02} --
\cite{galfil_84}).

\begin{figure}
 \centering
\includegraphics[height=60mm]{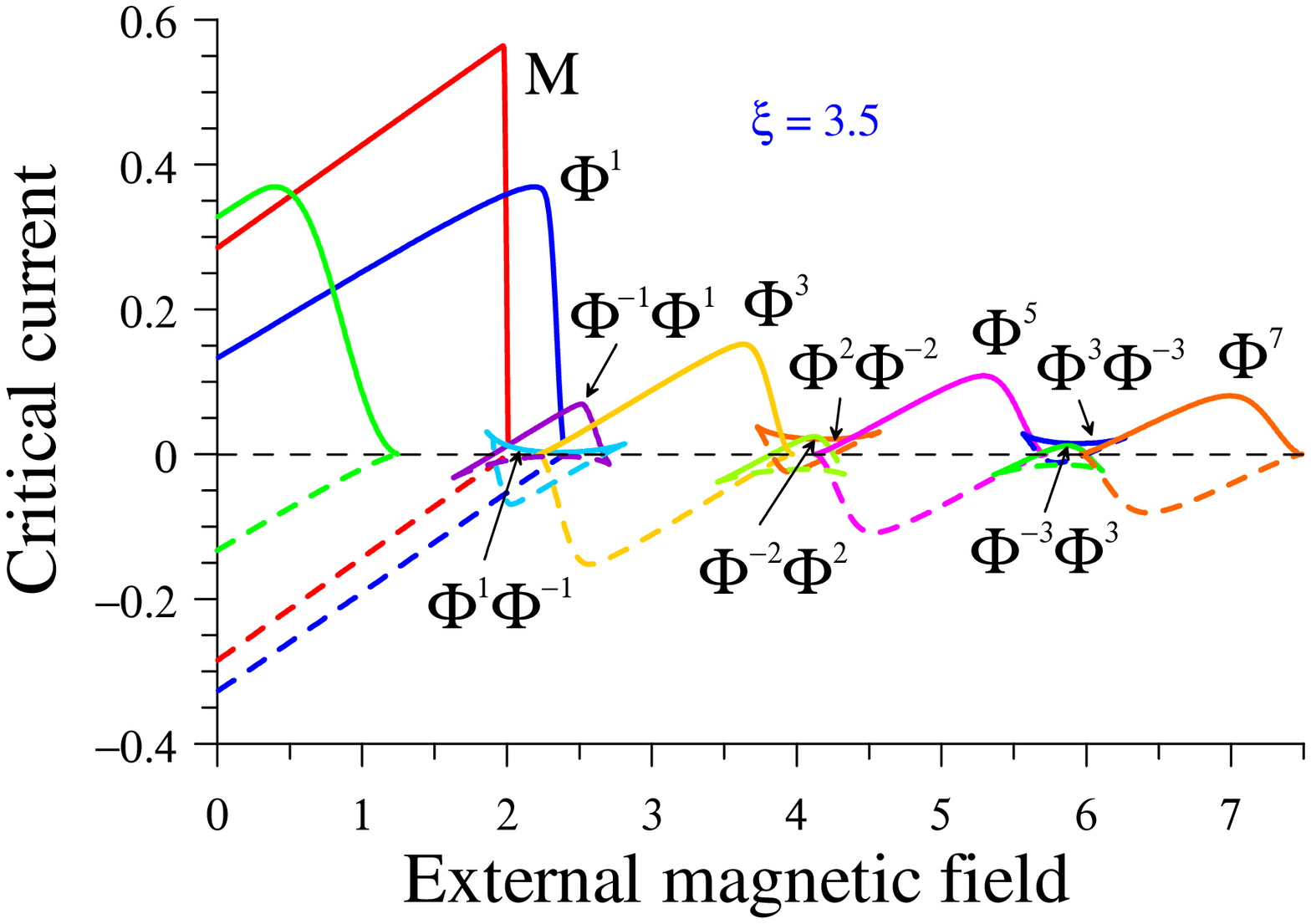} \includegraphics[height=60mm]{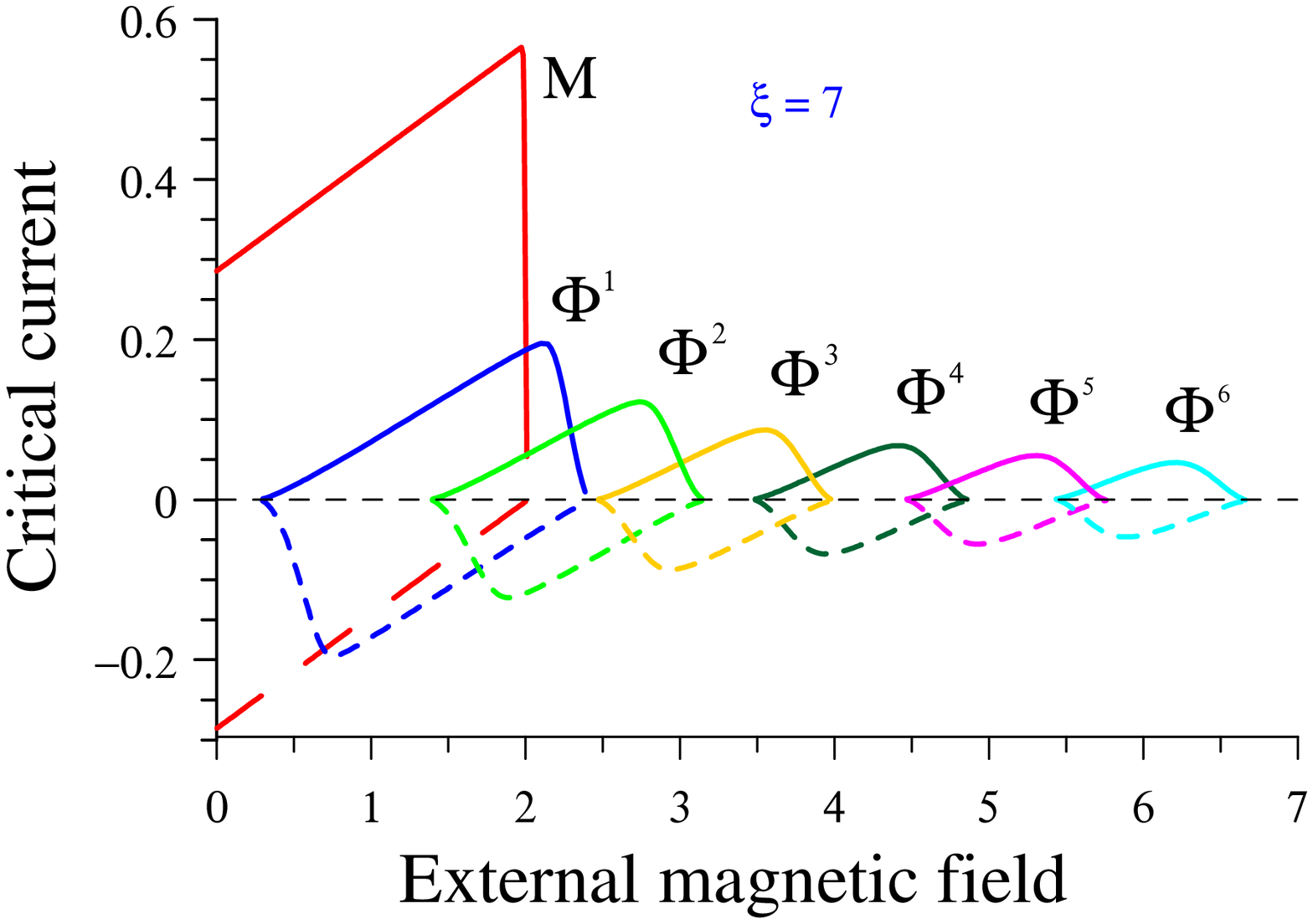}
\caption{Critical curve of the junction with the inhomogeneity at the center of junction
(left) and at the end of the junction(right). The length of the junction is $L=7$ and
width of inhomogeneity is $\Delta=0.7$} \label{in3.5}
\end{figure}

Results of numerical solution of non-linear eigenvalue problem \eqref{inline}  in the
in-line geometry is presented in Fig.\ \ref{gcrh_l7o}, where we demonstrate an influence
of the inhomogeneity position on the bifurcation curves ``critical current -- external
magnetic field''  for $L=7$ and $\Delta=0.7$. As we can see, the inhomogeneity in the
center of JJ leads to the appearance of stable mixed states like $\Phi^n\Phi^{m}$ ($n, m
= \pm 1, \pm 2, \ldots$, $n \ne m$ and $n m < 0$) and to the non-monotonic decrease of
maximums of critical curves for pure fluxon states with magnetic field. Here $\Phi^{\pm
1}$ denotes single fluxon (antifluxon) state in the junction.  In Fig.\ \ref{gcrh_l7o}
(right) we demonstrate a disappearance of mixed fluxon-antifluxon states when the
position of inhomogeneity is shifted to the end of the junction. The same result we
observed in the overlap geometry.

\begin{figure}
\begin{center}
\epsfxsize=0.7\hsize \epsfig{figure=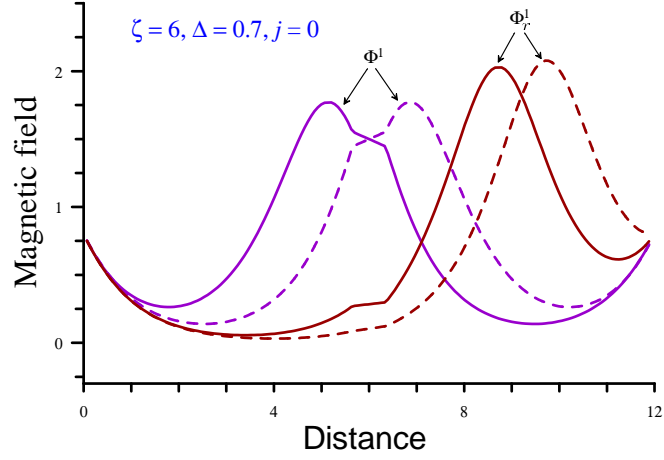,height=6cm}
\caption{The distribution of the magnetic field along the junction for $\Phi^1$ and
$\Phi^1_r$ at $h_e=0.8$ and $\gamma = 0$} \label{distance_l12_1inhom_ksi6_He08}
\end{center}
\end{figure}

In long JJ with the inhomogeneity in the center of junction in addition to the central
fluxon states appear the asymmetric fluxon states. In the overlap geometry for the
junction with $L=12$ , width of inhomogeneity  $\Delta=0.7$ and the amplitude of
Josephson current through inhomogeneity $j_{C}=0$ we have observed the "left" and the
"right" fluxon states $\Phi^1_l$, $\Phi^1_r$, $(\Phi^2 \Phi^{-2})_l$ and $(\Phi^{-2}
\Phi^{2})_r$. The distribution of the magnetic field along the junction for $\Phi^1$ and
$\Phi^1_r$ is shown in Fig.\ \ref{distance_l12_1inhom_ksi6_He08}.

    \begin{figure}
    \centering
    \includegraphics[height=60mm]{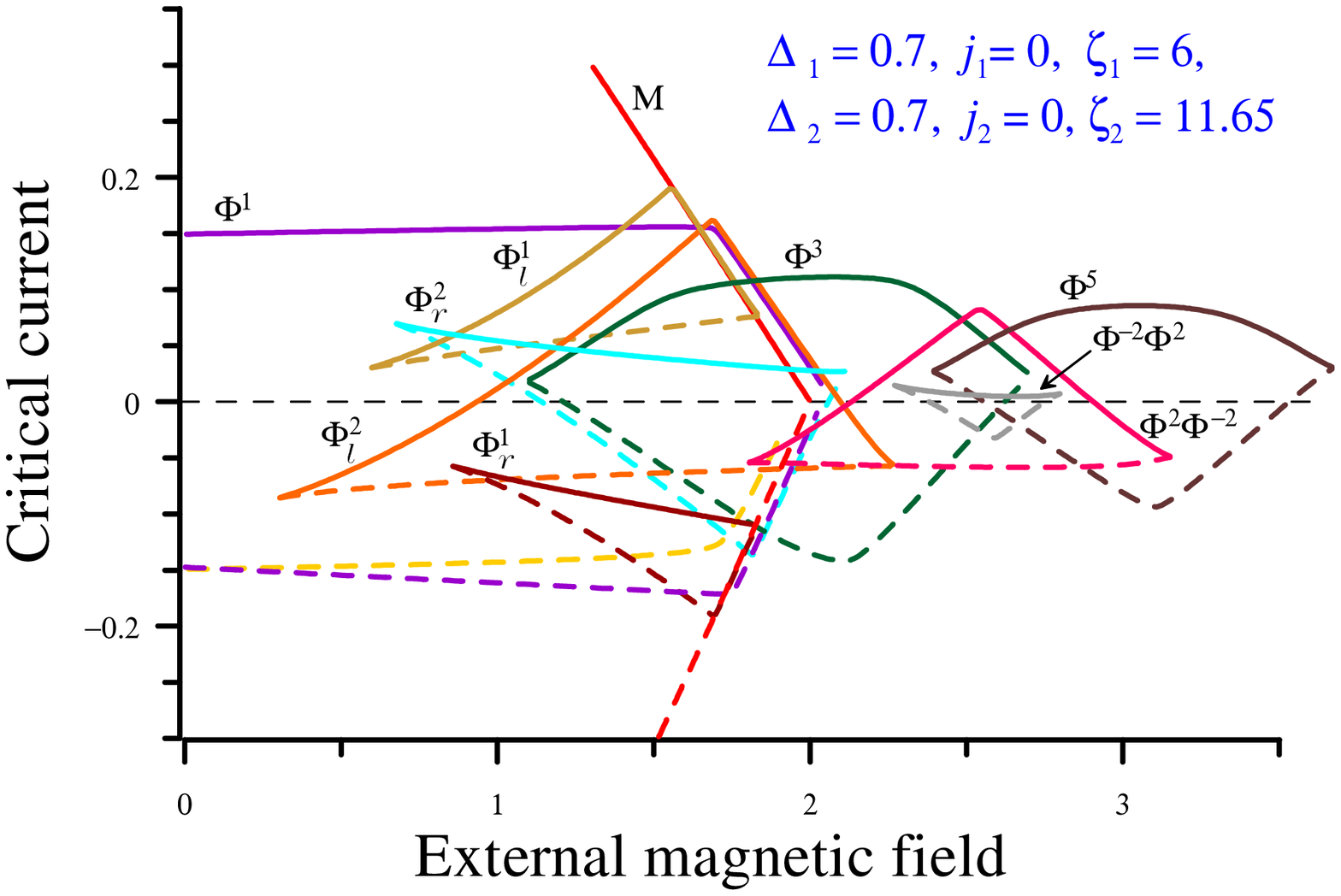}
    \includegraphics[height=60mm]{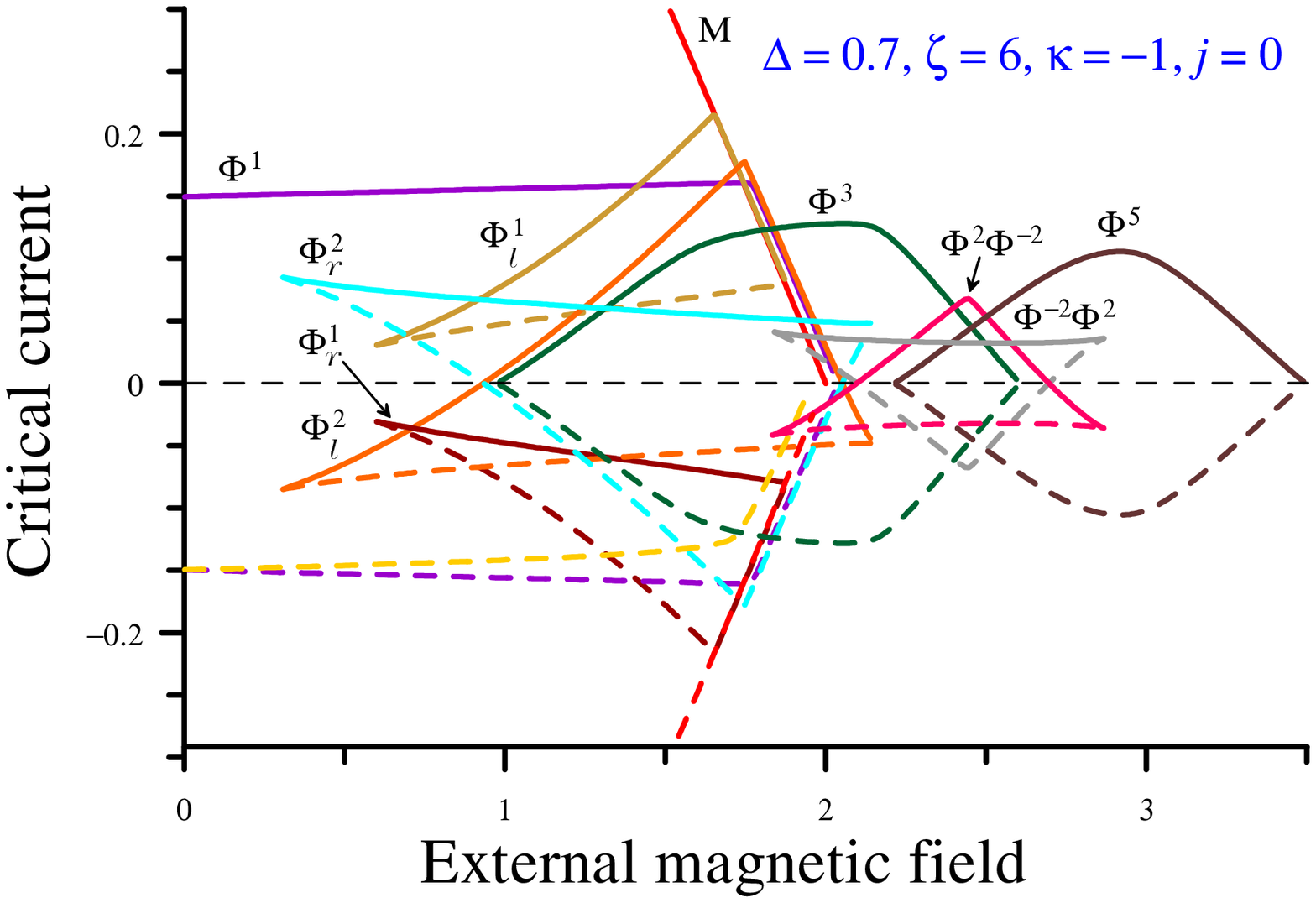}
     \caption{Critical curves for the junction with one inhomogeneity at the
center with $j_{C1}=0$ and another one at the end of the junction with $j_{C2}=1$ (a)
and $j_{C2}=0$ (b) . The length of the junction is $L=12$ and width of inhomogeneities
are $\Delta=0.7$} \label{grch_l12_2inhom_ksi6_j0_ksi1165_j0}
\end{figure}

We compare the bifurcation curves for the junction with one inhomogeneity at the center
with a case of JJ with two resistive inhomogeneities. The position of the first one was
fixed in the center and second one at the right end of the junction. The result of
simulation for $\Delta_1=\Delta_2=0.7$ and  $j_{C1}=j_{C2}=0$ is presented in Fig.\
\ref{grch_l12_2inhom_ksi6_j0_ksi1165_j0}.  We study the influence of the value $j_{C2}$
on the bifurcation curves of the junction and get the next main results. The decrease of
the current through the inhomogeneity on the right side of the junction makes a weak
influence on the curves $\Phi^1_l$ and $\Phi^2_l$ but essentially changes the curves
$\Phi^1_r$ and $\Phi^2_r$. There is strong effect on the bifurcation curves of mixed
states: stability region for $\Phi^2 \Phi^{-2}$ is increased, and for $\Phi^{-2}
\Phi^{2}$ is decreased. We observe that this decrease in $j_{C2}$ smoothes the maximums
on the dependence "critical current - magnetic field".

The bifurcation curves for mixed states demonstrate an interesting peculiarity. In some
intervals of magnetic field these states are stable only if the current through junction
is not equal to zero. We call such intervals as CC-regions (created by current). When
the inhomogeneity is shifted from the center of the junction, the pure fluxon states
have  the CC-regions as well. In our case with two inhomogeneities we have observed the
appearance of CC-regions  for pure fluxon states by the decrease of the current through
the second inhomogeneity. We found also that $\Phi^1_l$ and $\Phi^1_r$ are not stable
without current at all.



\begin{thebibliography}{10}

\bibitem{tlb_02} T.L. Boyadjiev, Numerical investigation of some critical regimes in non-linear
physical field models, DsC dissertation, Dubna, 2002.

\bibitem{bpp_jinr88}  T.L. Boyadjiev et al,  Comm. JINR P11-88-409, Dubna, 1988;



\bibitem{sbs_05} E.G. Semerdjieva, T.L. Boyadjiev, and Yu.M. Shukrinov, FNT, {\bf 31}, 1110 (2005).


\bibitem{galfil_84}Yu.S. Gal'pern and A.T. Filippov, Sov. Phys. JETP, \textbf{59}, 894 (1984).


\end{thebibliography}
\end{document}